\documentclass[10pt,letterpaper]{article}
\usepackage{opex3}
\usepackage[mathscr]{eucal}
\usepackage{amsmath}
\usepackage{amssymb}
\usepackage{amsfonts}
\usepackage{ae}
\usepackage{epsfig}
\usepackage{subfigure}
\usepackage{caption}
\usepackage{graphicx}
\usepackage{dcolumn}
\usepackage{bm}
\usepackage[latin1]{inputenc}
\usepackage{color}

\begin{document}
\title{Direct generation of a multi-transverse mode non-classical state of light}

\author{Benoît Chalopin$^{1,2}$, Francesco Scazza$^{1}$, Claude Fabre$^{1}$ and Nicolas Treps$^{1}$}

\address{$^{1}$ Laboratoire Kastler Brossel, Université Pierre et Marie Curie, ENS, CNRS \\ 4 place Jussieu, 75005 Paris, France}
\address{$^{2}$ Max Planck Institute for the Science of Light, Staudtstrasse 7/B2, 91058 Erlangen, Germany}

\begin{abstract}
  Quantum computation and communication protocols require quantum resources which are in the continuous variable regime squeezed and/or quadrature entangled optical modes. To perform more and more complex and robust protocols, one needs sources that can produce in a controlled way highly multimode quantum states of light. One possibility is to mix different single mode quantum resources. Another is to directly use a multimode device, either in the spatial or in the frequency domain. We present here the first experimental demonstration of a device capable of producing simultanuously several squeezed transverse modes of the same frequency and which is potentially scalable. We show that this device, which is an Optical Parametric Oscillator using a self-imaging cavity, produces a multimode quantum resource made of three squeezed transverse modes.
\end{abstract}


As the complexity of quantum communication and computation protocols increases, the need for quantum systems of high dimensionality increases accordingly. As far as light is concerned, this means that one must generate appropriately tailored  multimode non-classical light, the different modes being either frequency modes  \cite{Nussenzveig,Pfister,Silberhorn} and/or spatial modes  \cite{Ottavia,Blanchet,Janousek}. In particular, in the Continuous Variable regime, transport and processing of quantum information requires the use of  quantum multimode quadrature-entangled light beams, which can be produced by the linear mixing of squeezed states in different optical modes \cite{ReidEPR}.   So far, the complexity of multimode squeezing and entanglement experiments has increased to up to four modes \cite{Furusawa} with application to cluster-state based quantum computation. The quantum properties of a light beam consisting of several copropagating transverse modes of a monochromatic laser beam has recently raised a strong interest, both theoretically \cite{Kolobov,Lopez} and experimentally \cite{Janousek,Lassen2009,Manip01}. We present in this paper a source for multimode squeezing of three transverse modes, which, according to the theoretical model developped in \cite{Lopez}, can be scalable to a much larger number of modes.\\

\begin{figure}%
\begin{center}
\includegraphics[width=0.7\columnwidth]{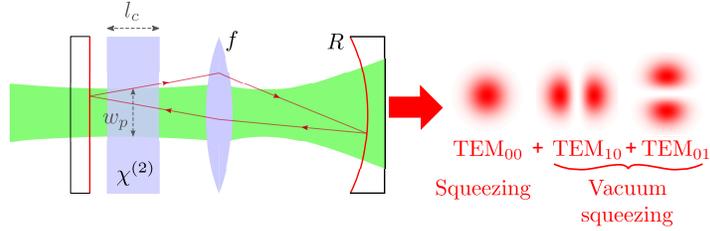}%
\caption{Sketch of the self-imaging OPO. The cavity contains a plane mirror, a lens and a spherical mirror, and reaches complete transverse mode degeneracy for precise values of their relative distances. A non-linear crystal is placed inside the cavity. The parametric downconversion process of pump photons generate  degenerate signal and idler photons in different spatial modes. The output of the OPO is a set of squeezed transverse modes.}%
\label{cavite}%
\end{center}
\end{figure}

Optical parametric amplificators (OPA) and oscillators (OPO) are amongst the most reliable and efficient sources for generating non-classical beams \cite{Kimble,Schnabel}. They use parametric down-conversion of an intense pump laser beam in a optical cavity. The photons created in pairs are strongly correlated, both temporally and spatially,  but the single-mode cavity usually loses all the transverse correlations. The experiment we present here relies on the use of a multimode cavity, the self-imaging cavity. Introduced by Arnaud \cite{Arnaud}, this cavity allows the simultaneous resonance of several transverse modes of a monochromatic laser beam. This enables the cavity build-up of images instead of just Gaussian beams. It has been used for example to efficiently frequency double images of various shapes \cite{DoublageImageant}. The cavity, depicted in Fig. \ref{cavite} consists of a plane mirror, a converging lens of focal lens $f$ and a spherical mirror of radius of curvature $R$. The cavity is self-imaging when the distances $L_1$ and $L_2$ between the optical elements verify the relations: $L_1=L_{1,deg}=f+f^2/R$ and $L_2=L_{2,deg}=f+R$, in which case the cavity is completely degenerate and allows the resonance of many transverse mode of a monochromatic laser beam, with a limit set by the transverse size and the aberration of the optical elements. If these conditions are not exactly fulfilled, one can still have the resonance of several transverse modes, as described in \cite{DoublageImageant}. Placing a $\chi^{(2)}$ nonlinear crystal inside this cavity, a self-imaging OPO is built, as proposed in \cite{Lopez}. \\

Since the cavity is completely degenerate, one can choose any mode basis to describe the transverse modes resonating inside the cavity \cite{DoublageImageant}. In \cite{Lopez}, it was showed that the best basis to describe the non-classical states produced by a self-imaging OPO is the eigenbasis of the parametric down-conversion process of the corresponding pump profile, which can be found through the diagonalization of the coupling matrix. This matrix is given by the product between the phase-matching coefficient, the transverse profile of the pump and the longitudinal overlap between modes through diffraction. In many practical cases, this eigenbasis is very close to a set of Hermite-Gauss polynomials. Assuming equal loss rate $\gamma$ for all modes, the evolution of the mode bosonic operator $\hat{S}_k$ associated to each eigenmode is given by:
\begin{equation}
\frac{d \hat{S}_k}{dt}= -\gamma \hat{S}_k + \Lambda_k \hat{S}_k^\dagger + \sqrt{2 \gamma} \hat{S}_{k,in}
\end{equation} 
where $\Lambda_k$ is proportionnal to the eigenvalue associated to each eigenmode, and $\hat{S}_{k,in}$ the input operator for each mode. The output state of the self-imaging OPO is a then a set of independent squeezed modes, each associated to a pump amplitude proportionnal to $\Lambda_k$. Hence, eigenvalues $\Lambda_k$ determine the maximum amount of squeezing of each eigenmode $k$: 
\begin{equation}
V_{k,min}=\frac{|\Lambda_0|-|\Lambda_k|}{|\Lambda_0|+|\Lambda_k|}
\end{equation}
where $\lambda_0$ is the eigenvalue of largest modulus. The spectrum of the coupling matrix exhibits therefore the multimode non-classical capabilities of the self-imaging OPO: in particular, one sees that a highly multimode nonclassical  beam is generated when many $|\Lambda_k|$ are nonzero and close to $|\Lambda_0$.\\

The analysis of the coupling matrix in our present configuration show that, to generate several non-classical modes, the waist of the pump beam of the OPO must be larger than the crystal coherence length $l_{coh}=\sqrt{\lambda l_c/\pi n}$, as defined in \cite{Lopez}, where $\lambda$ is the wavelength of the signal field, $l_c$ the crystal length, and $n_s$ its refractive index at $\lambda$. For a 10-mm long PPKTP crystal, we find $l_{coh} \approx$ 40 $\mu$m. With a pump waist larger than this value, one gets multimode squeezing at the output of the device. Moreover, if the pump profile is gaussian, the eigenmodes are very close to being a set of modes with orthogonal Hermite-Gauss profiles. The expected number of modes can be calculated using the cooperativity parameter \cite{Walmsley}, which depends quadratically on the waist of the pump. Note that the OPO threshold is also a quadratic function of the pump waist, so that the number of generated squeezed modes is only limited by the available pump power.\\
\begin{figure}%
\begin{center}
\includegraphics[width=0.7\columnwidth]{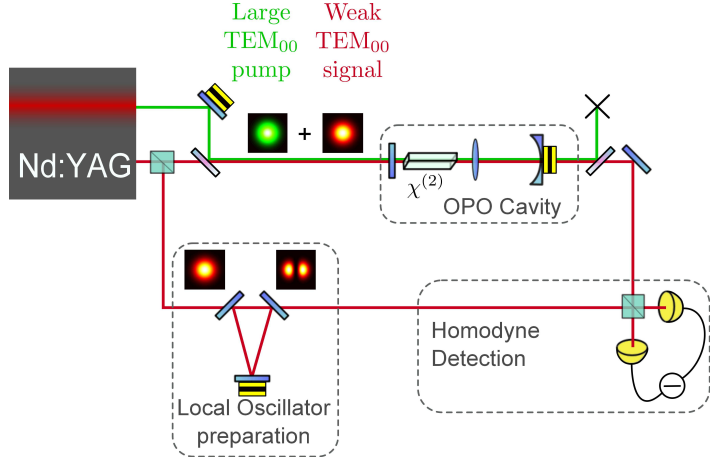}%
\caption{Sketch of the experimental setup.}%
\label{expsetup}%
\end{center}
\end{figure}

The experimental setup is depicted on Fig. \ref{expsetup}. Inside the self-imaging cavity, the spherical mirror has a radius of curvature $R=$ 50 mm, and the lens a focal length $f=$ 30 mm, which gives a total cavity length of 128 mm at degeneracy. The cavity finesse is $\mathcal{F}=$ 250 which leads to a cavity bandwidth of 4.7 MHz. Finally the escape efficiency is around 60 \%, mainly because of the extra losses on the intra-cavity lens. The highly nonlinear efficiency of the periodically poled potassium titanyl phosphate (PPKTP, 10 mm long, dual anti-reflexion coated) enables us to keep reasonable non-linear efficiency with a single-pass 532-nm pump beam from a frequency-doubled 1064-nm yttrium aluminium garnet (YAG) laser. The pump waist, about 120 $\mu$m in our case, is chosen in order to obtain the maximum number of non-classical modes compatible with the pump power at our disposal. In our current setup its theoretical value is 7, and all the theoretical eigenmodes have an overlap above 99.5 \% to the corresponding Hermite-Gauss functions.  In order to lock the cavity, and later align the homodyne detection, we seed the cavity with a gaussian beam whose waist size  $w_0=$ 90$\mu$m is so that it is adjusted to the first eigenmode of the OPO and is positioned in the center of the PPKTP crystal, near the plane mirror of the cavity, as described in Fig. \ref{cavite}. The cavity is locked at resonance using a Pound-Drever-Hall locking, and the relative phase between the pump and the seed is locked using an non-linear interferometer that uses second harmonic generation of part of the seed, as described in \cite{Manip01}.\\

\begin{figure}%
\begin{center}
\includegraphics[width=0.7\columnwidth]{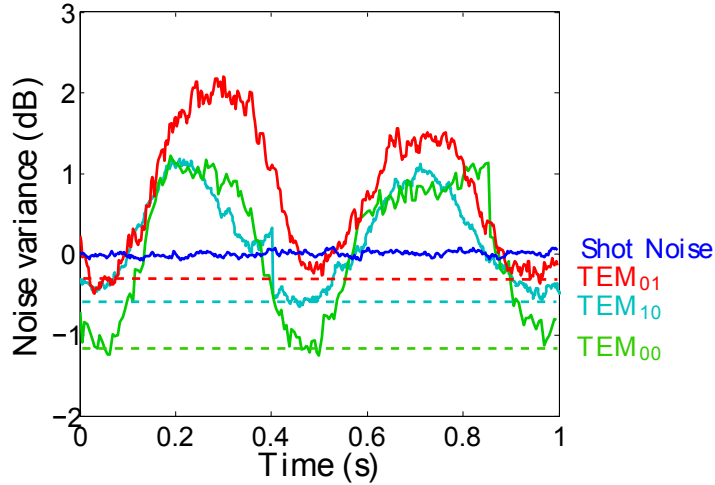}
\caption{a) Output of the self-imaging OPO, three transverse modes of the same wavelength are squeezed. b)
  Fluctuations of three orthogonal output modes of the OPO measured with a homodyne detection. The phase of the local oscillator is swept, so that the fluctuations are measured on all different quadratures. The blue line indicates the shot noise level.The TEM$_{00}$ (red curve) shows 1.2 dB of squeezing below shot noise, the TEM$_{10}$ (cyan curve) 0.6 dB and the TEM$_{01}$ (green curve) 0.3 dB.}%
\label{sqz}%
\end{center}
\end{figure}

The quantum fluctuations of the different modes contained in the output state of the OPO are measured  with a homodyne detection whose local oscillator can be switched from one mode to another through a cavity mode-converter \cite{Lassen2006}. Unlike other experiments \cite{Lassen2006}, the OPO cavity remains locked on the same resonance while the mode converter is switched from one mode to another. The alignment procedure is the following: we first mode-match the local oscillator to the transmitted seed, when there is no pump beam in the cavity. The local oscillator is filtered by a ring-cavity mode cleaner, as shown on Fig.\ref{expsetup}. Changing the resonance of this cavity from one transverse mode to another enables us to change the transverse mode of the local oscillator, thus changing the output mode analyzed by the homodyne detection, but without changing the alignment. This feature ensures that the different modes we analyzed with the homodyne detection are orthogonal.\\
The results are presented on Fig.\ref{sqz}. We observe that at least three copropagating transverse modes have fluctuations below shot noise, with 1.2 dB of squeezing for the TEM$_{00}$, 0.6 dB for the TEM$_{10}$ and 0.3 dB for the TEM$_{01}$. These fluctuations were measured at a frequency of 3 MHz. The higher order modes were analyzed, but the amount of squeezing was too low to clearly distinguish it from the shot noise. \\

The amount of squeezing we measure is in agreement with theoretical expectations taking into account all different parameters of our experiment, which cannot all be optimized for maximum squeezing. For example the lens inside the cavity added significant losses, which decreased greatly the escape efficiency of the system and therefore the maximum amount of squeezing that can be reached in our set-up . We compared our multimode OPO with an equivalent single-mode OPO by simply slightly decreasing the cavity length. We found then that the pure TEM$_{00}$, single mode output of the OPO had the same squeezing as the most squeezed mode of the multimode OPO, which is in agreement with the theoretical prediction \cite{Lopez}.

The squeezing of the two other modes can also be compared to the theory. If we set the losses of the system at a value where the maximum squeezing is 1.2 dB, given the size of the pump beam and the length of the crystal, the squeezing on both these modes should be 0.9 dB. The difference with our measurement can be explained by the fact that the cavity is not completely degenerate, and that a slight cavity detuning is present on both of these modes. Finally, the difference between the TEM$_{10}$ and the TEM$_{01}$ can be explained by a non-homogenous transverse phase shift in the non-linear crystal due to the periodic poling itself, which is done by applying a large voltage between two facets of the crystal. This obviously introduces a limitation to the multimode performances of our system. However, several technical aspects can be improved. For example, with higher available pump power, one can decrease the cavity finesse and therefore increasing the escape efficiency and the cavity degeneracy, without increasing the thermal lens effects, because the pump beam is taken with a larger waist . One may also use another type of crystal, non-periodically poled, which would enable more degeneracy between the two transverse axes.\\

To conclude, we have presented an experiment that showed for the first time the direct generation of a three-mode squeezed state, on three well-defined orthogonal transverse modes. The results can also be interpreted as a verification of the theory developped in \cite{Lopez}, which stated that the self-imaging OPO can produce a set of squeezed copropagating transverse modes. Moreover, these modes are Hermite-Gauss modes, and their number and size can be engineered through the crystal and pump parameters, which makes this self-imaging OPO a potentially scalable source for multimode squeezing and entanglement.

We acknowledge the financial support of the Future and Emerging Technologies (FET) programme within the seventh Framework Programme for Research of the European Commission, under the FET-Open grant agreement HIDEAS FP7-ICT-221906.


\begin{thebibliography}{99}
\bibitem{Braunstein} S. L. Braunstein and P. van Loock, "Quantum information with continuous variables", Rev. Mod. Phys. \textbf{77}, 513 (2005) 
\bibitem{Lopez} L. Lopez, B. Chalopin, A. Rivière de la Souch\`ere, C. Fabre, A. Ma\^itre, and N. Treps, "Multimode quantum properties of a self-imaging optical parametric oscillator: Squeezed vacuum and Einstein-Podolsky-Rosen-beams generation", Phys. Rev. A \textbf{80}, 043816 (2009)
\bibitem{Nussenzveig} A. S. Coelho, F. A. S. Barbosa, K. N. Cassemiro, A. S. Villar, M. Martinelli and P. Nussenzveig, "Three-color entanglement", Science \text{326} 823 (2009)
\bibitem{Pfister} N. C. Menicucci, S. T. Flammia and O. Pfister, "One-Way Quantum Computing in the Optical Frequency Comb", Phys. Rev. Lett. \textbf{101} 130501 (2008)
\bibitem{Silberhorn} A. Eckstein and C. Silberhorn, "Broadband frequency mode entanglement in waveguided parametric downconversion", Optics Letters, \textbf{33} 1825 (2008)
\bibitem{Ottavia} O. Jedrkiewicz, Y-K. Jiang, E. Brambilla, A. Gatti, M. Bache, L. A. Lugiato and P. Di Trapani, "Detection of Sub-Shot-Noise Spatial Correlation in High-Gain Parametric Down Conversion", Phys. Rev. Lett \textbf{93} 243601 (2004)
\bibitem{Blanchet} J-L. Blanchet, F. Devaux, L. Furfaro and E. Lantz, "Measurement of Sub-Shot-Noise Correlations of Spatial Fluctuations in the Photon-Counting Regime", Phys. Rev. Lett \textbf{101} 233604 (2008)
\bibitem{Janousek} J. Janousek, K. Wagner, J.F. Morizur, N. Treps, P.K. Lam, C.C. Harb, and H.A. Bachor, "Optical entanglement of co-propagating modes", Nature Photonics \textbf{3} 399 (2009)
\bibitem{ReidEPR} M. D. Reid, "Demonstration of the Einstein-Podolsky-Rosen paradox using nondegenerate parametric amplification", Phys. Rev. A \textbf{40} 913 (1989)
\bibitem{Furusawa} M. Yukawa, R. Ukai, P. van Loock and A. Furusawa, "Experimental generation of four-mode continuous-variable cluster states", Phys. Rev. A \textbf{78} 012301 (2008)
\bibitem{Kolobov} M. I. Kolobov, "The spatial behavior of nonclassical light", Rev. Mod. Phys. \textbf{71} 1539 (1999)
\bibitem{Lassen2009} M. Lassen, G. Leuchs, and U.L. Andersen, "Continuous Variable Entanglement and Squeezing of Orbital Angular Momentum States", Phys. Rev. Lett. \textbf{102} 163602 (2009)
\bibitem{Manip01} B. Chalopin, F. Scazza, C. Fabre and N. Treps, "Multimode nonclassical light generation through the optical-parametric-oscillator threshold", Phys. Rev. A \textbf{81} 061804 (R) (2010)
\bibitem{Kimble} L-A. Wu, H. J. Kimble, J. L. Hall and H. Wu, "Generation of squeezed states by parametric down conversion", Phys. Rev. Lett. \textbf{57} 2520 (1986)
\bibitem{Schnabel} T. Eberle, S. Steinlechner, J. Bauchrowitz, V. H\"andchen, H. Vahlbruch, M. Mehmet, H. M\"uller-Ebhardt, and R. Schnabel, "Quantum Enhancement of the Zero-Area Sagnac Interferometer Topology for Gravitational Wave Detection", Phys. Rev. Lett. \textbf{104} 251102 (2010)
\bibitem{Arnaud} J.A. Arnaud, "Degenerate optical cavities", Applied Optics, Vol \textbf{8}. Issue 1, page 189 (1969)
\bibitem{DoublageImageant} B. Chalopin, A. Chiummo, C. Fabre, A. Maitre and N. Treps, "Frequency doubling of low power images using a self-imaging cavity", Opt. Exp. \textbf{18} 8033 (2010)
\bibitem{Walmsley} C.K. Law, I.A. Walmsley and J.H. Eberly, "Continuous Frequency Entanglement: Effective Finite Hilbert Space and Entropy Control", Phys. Rev. Lett. \textbf{84} 5304 (2000)
\bibitem{Lassen2006} M. Lassen, V. Delaubert, C. Harb, P.K. Lam, N. Treps, and H.A. Bachor, "Generation of squeezing in higher order Hermite-Gaussian modes with an optical parametric amplifier", JEOS RP \textbf{1} 06003 (2006)
\end{thebibliography}
\end{document}